\documentclass[a4paper,11pt]{article}
\usepackage{aaskaiid}
\usepackage{orcidlink}
\newcommand{\change}[1]{ #1  }

\title{Unveiling the Mysteries of Lightning: Exploring its fundamental Physical Processes with SKA-LOW}
\ShortTitle{Lightning}

\ShortName{Brian Hare et al.} % shortened name list for header 
\author[1,2]{B.M.~Hare \orcidlink{0000-0001-5138-1235}}
\author[3]{S.~Bouma\orcidlink{0000-0002-6959-2302}}

\author[4]{S.~Buitink\orcidlink{0000-0002-6177-497X}}

\author[4,5]{A.~Corstanje\orcidlink{0000-0001-5992-6228}}

\author[6]{S.~Cummer\orcidlink{0000-0002-0002-0613}}

\author[7]{J.~Dwyer\orcidlink{0000-0002-3581-1503}}

\author[4]{V.~De~Henau\orcidlink{0009-0003-0337-3558}}

\author[8,4]{T.~Huege\orcidlink{0000-0002-2783-4772}}

\author[3]{P.~Laub\orcidlink{0009-0003-2617-9109}}

\author[7]{N.~Liu\orcidlink{0000-0002-8626-6218} }

\author[1,2]{M.A.A.~Lourens\orcidlink{0009-0006-3640-1043} }

\author[5, 9]{K.~Mulrey\orcidlink{0000-0001-8026-8020}}

\author[4,10]{A.~Nelles\orcidlink{0000-0002-1720-6350}}

\author[1,2]{O.~Scholten\orcidlink{0000-0003-3649-1254}}

\author[1,2]{C.~Sterpka\orcidlink{0000-0001-8217-0836} }

\author[4]{K.~Terveer\orcidlink{0009-0002-9594-0419}}

\author[1,2]{P.~Ťureková\orcidlink{0009-0006-1262-7507}}

\author[8]{K.~Watanabe\orcidlink{0000-0003-0599-4035}}

\affiliation[1]{Netherlands Institute for Radio Astronomy (ASTRON), Dwingeloo, Netherlands} 

\affiliation[2]{Kapteyn Astronomical Institute, University Groningen, Groningen, Netherlands} 

\affiliation[3]{Erlangen Centre for Astroparticle Physics (ECAP), Friedrich-Alexander-University Erlangen-N\"urnberg, 91058 Erlangen, Germany 
} 

\affiliation[4]{Inter-University Institute For High Energies (IIHE), Vrije Universiteit Brussel (VUB), Pleinlaan 2, 1050 Brussels, Belgium
} 

\affiliation[5]{Department of Astrophysics/IMAPP, Radboud University Nijmegen, Nijmegen, Netherlands} 

\affiliation[6]{Electrical and Computer Engineering Department, Duke University, Durham, NC, USA}

\affiliation[7]{Department of Physics and Astronomy \& Space Science Center (EOS), University of New Hampshire Durham, New Hampshire, USA}

\affiliation[8]{Institut f\"ur Astroteilchenphysik, Karlsruhe Institute of Technology (KIT), P.O.~Box 3640, 76021 Karlsruhe, Germany
}

\affiliation[9]{Nationaal Instituut voor Kernfysica en Hoge Energie Fysica (NIKHEF), Science Park, Amsterdam, Netherlands
}

\affiliation[10]{Deutsches Elektronen-Synchrotron DESY, Platanenallee 6, 15738 Zeuthen, Germany}

\emailAdd{hare@astron.nl}

\abstract{

Lightning is a surprisingly poorly understood phenomena. It consists of a wide variety of complex processes such as initiation, propagation, connection to ground, even emission of high-energy radiation. However, due to the extreme challenges in observing lightning at fast time scales, small spatial scales, and behind obscuring clouds, these processes are not well understood. In the past, interferometers such as the LOFAR radio telescope have provided unique insight and discoveries into the physics of lightning. The new SKA-LOW being built in western Australia will provide unrivaled spectral bandwidth and sensitivity, which will be combined with high resolution resulting from large antenna baselines. We will use SKA-LOW to observe lightning in order to explore its fundamental plasma physics, such as how it initiates and propagates. SKA's high bandwidth will allow us to test how lightning emits VHF radiation, giving tremendous insight into precisely how the plasma behaves. SKA's sensitivity will allow us to explore extremely faint lightning processes, such as the very first radio emission from a lightning flash. Here, we detail the lightning physics that can be explored with SKA, as well as the observation strategy needed explore such physics.
}

\begin{document}
\maketitle
% CITE -> citep

\section{Introduction}

\begin{figure}
\centering
	\includegraphics[width=0.95\linewidth]{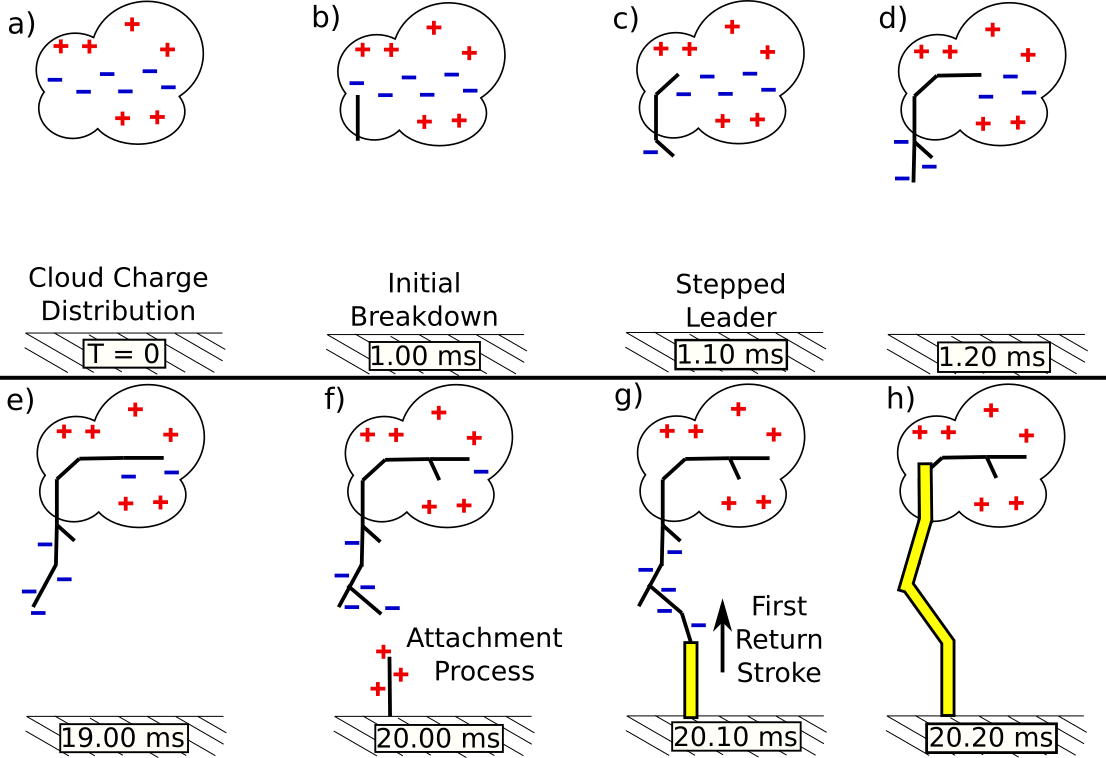}
	\caption{\label{fig:LightningPhenom} Some basic lightning processes. Starting with a) the distribution of electrical charge in a thunderstorm, followed by b) initiation, c-e) propagation of the leaders, f) connection to ground, and g-h) return stroke. Inspired by figure 1.3 in \cite{Dwyer:2014}. }
\end{figure}

Lightning occurs so rapidly, ionizing tens of kilometers of air in a split second, that casual observers can be forgiven for thinking that it happens all at once as a single process. The truth, however, is that despite occurring so quickly lightning consists of a bewildering variety of different phenomena and processes. Figure \ref{fig:LightningPhenom} shows a few of the basic processes of a lightning flash. Lightning starts with initiation (panel b), where a small region of air (exact size unknown) undergoes dielectric breakdown and turns into a conducting plasma. How exactly this initiation occurs \change{involves multiple complex steps and} is not well understood. The small region of plasma quickly polarizes in the thunderstorm electric field and so creates a strong field enhancement at its tips. This allows the charged tips to then propagate and branch throughout the thunderstorm (panels c-e). We refer to the generated plasma channels as leaders. In the type of lightning flash depicted in Figure \ref{fig:LightningPhenom}, the positively charged leaders propagate through the negatively-charged thundercloud region, and the negatively charged leaders propagate towards ground with speeds around $10^5$~m/s. The physics of how these plasma channels grow and propagate is not understood. The negative leaders tend to grow in jumps around 10~m long (and thus are often referred to as stepped leaders) and they emit copious amounts of VHF radiation as they grow, but the positive leaders grow smoothly and emit very little VHF radiation. It is not understood why negative leaders emit VHF radiation, or why positive leaders should be so different in this respect. As the negative leaders approach the ground there is a connection process (panel f), which is also not understood. Once the negative leader connects to ground there is a very strong current pulse that propagates up the plasma channel at around $10^8$~m/s to equalize the electric potential between the ground the rest of the flash, like a short-circuit (panels g-h). This process, called a return-stroke, emits large amounts of energy in the forms of light, heat, sound, and low-frequency radio radiation ($<10$~MHz). \change{In addition to these basic processes}, there are many other complex lightning processes, such as current pulses that re-energize decaying channels, bursts of gamma rays that can saturate particle detectors in space, lightning that propagates up into the ionosphere, and far more. None of these are well understood. \citep{Dwyer:2014}

Lightning is poorly understood due to two major reasons. First, is that it is difficult to observe with any precision (especially initiation). The majority of the flash occurs inside opaque thunderclouds at random places and times. High speed cameras have provided significant insight into lightning physics (e.g. \cite{Kong:2008, Hill:2011, Stolzenburg:2013, Saba:2022}). However, they are limited in that they can only observe the portions of the lightning near the ground with high resolution \change{due to the opaqueness of thunderstorms}. It is only in the last ten years, with the explosion of lightning radio interferometry, that we've been able to seriously explore intracloud lightning (e.g. \cite{Stock:2014, Rison:2016}). The second reason that lightning is poorly understood is that the plasma physics is extremely complicated; it is highly non-equilibrium in a regime that is not well-explored (full atmospheric pressure, strong electric fields, and wide range of temperatures) \citep{Nijdam:2020}. 

Propagation of a lightning leader is facilitated by millions of streamers produced at its tip (a streamer is a self-sustaining ionization wave that can propagate in an electric field much lower than the breakdown field resulting in a thin channel of low temperature plasma, and is essentially the ``atom'' of many lightning processes). However, state-of-the-art models can only simulate a few streamers at a time. \change{Thus, multi-scale modeling of lightning growth including the very large number of streamers present at a leader tip is currently computationally infeasible.} Even long laboratory sparks can only be used as a loose analog for lightning plasma physics since we cannot make laboratory sparks at the large scales (10~m and larger) needed to simulate the plasma phenomena inside of lightning. The result is that there is a large family of plasma processes that can only be studied through natural terrestrial lightning observations. 
%The result is that there is a large gap between the largest phenomena that can be modeled and the smallest phenomena that can be well-observed. 

\begin{figure}
\centering
	\includegraphics[width=0.7\linewidth]{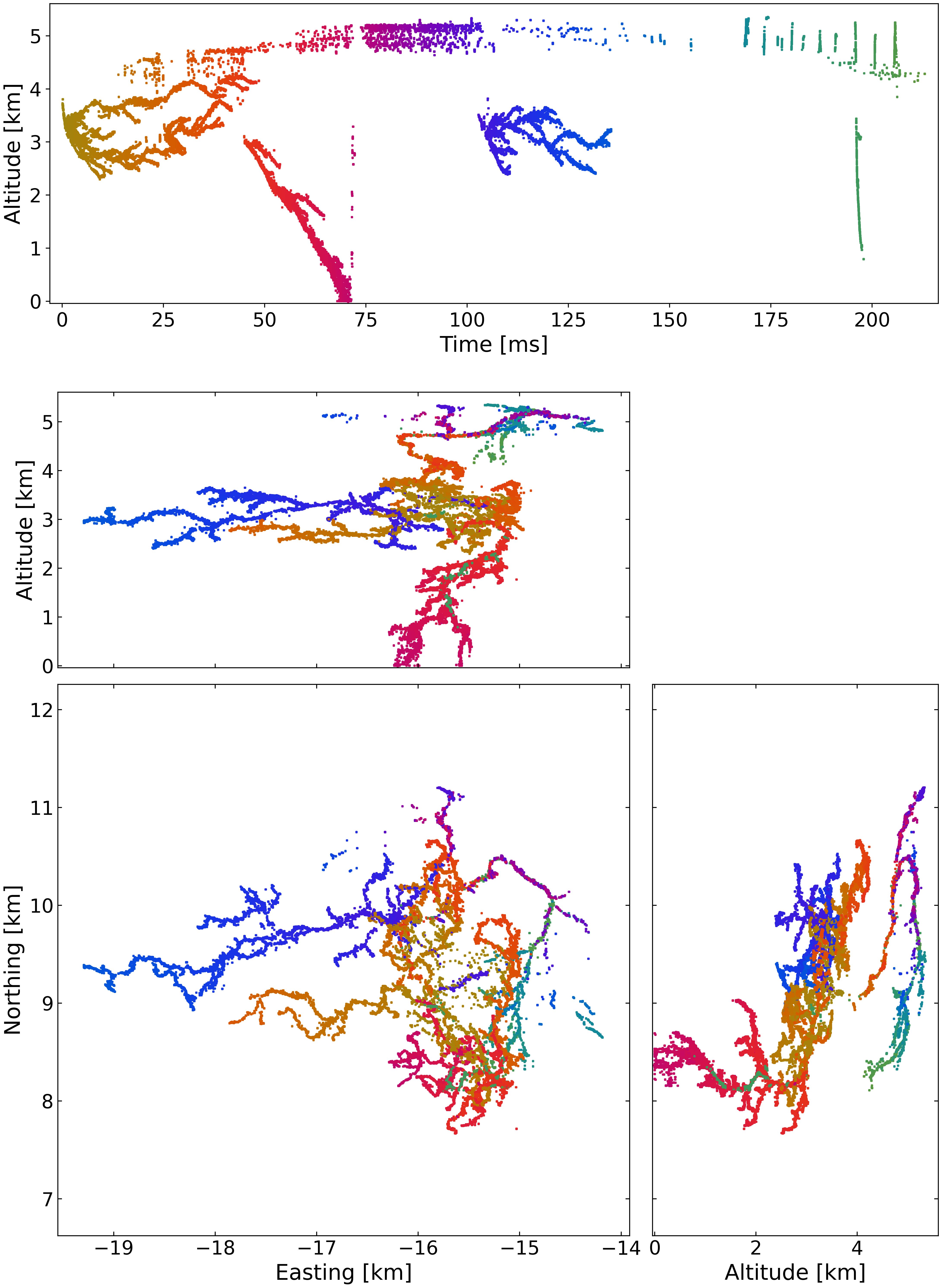}
	\caption{\label{fig:LofarFlash} A small lightning flash observed by LOFAR. Each dot is the 3D location and time of a located source of VHF emission during the lightning flash, projected onto different Euclidean planes. Color indicates time in order to help compare the different panels. }
\end{figure}

In the last decade lightning science has undergone a revolution. New phased array radio imaging instruments and techniques are finally allowing us to observe in-cloud lightning processes with high-precision \citep{Stock:2014}. Our group uses the LOFAR radio telescope \citep{Haarlem:2013} to image lightning with meter-scale and nanosecond-level resolution. Figure \ref{fig:LofarFlash} shows a small and relatively simple lightning flash imaged by LOFAR. Each dot is the 3D location and time of a located radio pulse-source during a lightning flash. Each panel of figure \ref{fig:LofarFlash}  shows the location and time of the radio source projected onto a 2D plane. Due to its wide/zoomed-out scale, most of the interesting physics in Figure \ref{fig:LofarFlash}  is shown in the altitude vs time panel. The lightning flash shown in figure \ref{fig:LofarFlash}  is only about 200~ms in duration and about 5~km long, wide, and tall. Typical lightning flashes are about 0.5~s in duration and 10~km in size. The flash in figure \ref{fig:LofarFlash}  starts at T=0, 3.5~km altitude. The actual initiation process is far too small (and weak) to see in this image. We do see the negative leaders propagate down slightly, and then propagate horizontally around 3~km altitude. At roughly T=40~ms the negative leaders begin to propagate down to ground. When the negative leaders connect with ground, at roughly T=65~ms, there is a return stroke which is seen as a vertical bar of radio sources in the altitude vs time panel (due to the high speed at which the return stroke propagates up the channel). The negative leaders start to propagate again at around T=100~ms, 3~km altitude. At the same time there is VHF emission from the positive leader above 4~km altitude, between T=25~ms to T=150~ms. This VHF emission is actually not due to the propagation of the positive leaders (which are very quiet in VHF), but is actually due to a lightning phenomena recently discovered by our group  \citep{Hare:2019}, we call needles. Which are a form of negative breakdown on the positive leader. Later in this flash, after T=160~ms, there are a series of vertical bars (in altitude vs time) of radio sources. Each of these vertical bars is called a dart leader which is a charge-pulse that propagates along previously-established plasma channels and re-heats the plasma channel. Since the decaying channels are still warm and somewhat-ionized, they are relatively easy to re-ionize, thus the dart leaders propagate very quickly (~$10^7$~m/s, thus their name).

%Lightning observations with LOFAR work very differently than astronomical observations, and actually have a similar strategy as cosmic ray air shower observations (as a result, the LOFAR/SKA cosmic ray and lightning groups collaborate closely). Since during a thunderstorm the telescope is not usable for astronomy science, when a thunderstorm approaches LOFAR ($\approx$~100~km), the telescope is put into a dedicated observation mode. In this mode, similar to cosmic ray observations, the raw antenna voltages directly out of the digitizers (before beamforming, channelization, or any other processing) are continuously recorded to RAM (transient buffers). Then, when a flash occurs (as detected by a long-range thunderstorm detection network or by a dedicated antenna), the transient buffers are frozen and read to disk. The saving process takes about 20~minutes, during which LOFAR cannot observe any more lightning. 

%After observation we image the lightning offline with our own group-designed and dedicated imaging pipeline. 

Figure \ref{fig:LofarFlash} and similar images have enabled LOFAR to produce many breakthroughs in the field of lightning science. For example by discovering new lightning phenomena such as ``needles'', which are a form of negatively charged propagation away from ostensibly positive plasma channels, thus changing how we think about lightning channel charging \citep{Hare:2019}. LOFAR has imaged cloud-top discharges in high resolution, thus revealing that they mostly consist of the same phenomena as normal large-scale lightning flashes \citep{Scholten:2023}. In addition LOFAR has produced the first images of normal lightning initiation \citep{Sterpka:2021}. However, there is still a tremendous amount about lightning that is not understood. SKA-LOWs' bandwidth and sensitivity, as well as location, will produce even more deep and precise probes of lightning plasma physics that are needed to understand how lightning initiates, propagates, and emits short VHF pulses. 

\section{Science Cases for Lightning Observations with SKA-LOW}

\subsection{ Source of VHF emission }

One of the big questions in lightning science is: what is the emission mechanism behind VHF radiation? It is strongly believed that VHF emission from lightning  comes from streamer activity instead of the hot-conducting core of the leader plasma channels. The general physics argumentation is that the hot leader cores are many kilometers long and behave somewhat like long conducting wires, and thus emit at frequencies well below the VHF range ( and these low-frequency emissions are quite well studied, e.g. \cite{Dwyer:2014} ), whereas streamers, which are somewhere around a meter \citep{Saba:2022} and smaller \citep{Nijdam:2020} in size, are exactly the right size for VHF emission. Understanding why lightning emits VHF is important not only for understanding the plasma physics, but also for interpreting interferometric images made by SKA and other experiments around the globe.  

\begin{figure}
\centering
	\includegraphics[width=0.95\linewidth]{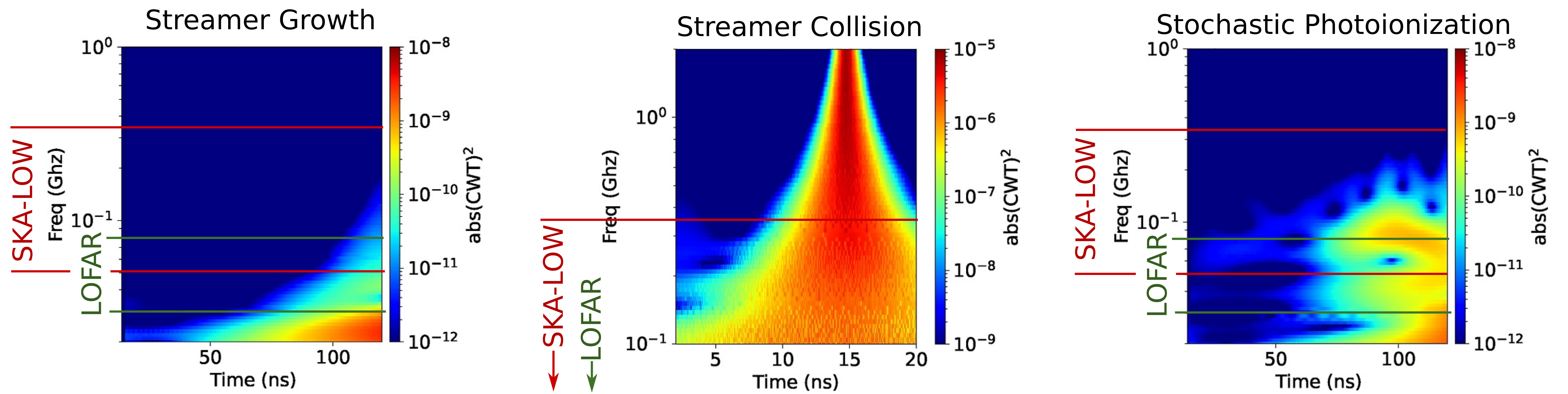}
	\caption{\label{fig:ModelSpectra} Frequency vs time plots of three VHF emission models: exponential streamer growth, streamer collision, and stochastic photo-ionization. Time is from start of the simulation. In panel "streamer collision", the two streamers collide at 14~ns. The frequency band of SKA-LOW and LOFAR-LBA is indicated. Color indicates power of the continuous wavelet transform on an arbitrary scale. Figure modified from \citep{Malla:2024}}
\end{figure}

Experimental evidence that VHF is emitted from streamers (as opposed to leader cores) starts with the fact that VHF emission comes from the tip of negative leader channels (where there are many active streamers) instead of from the body of the leader. In addition, it has been observed that each ``jump'' of a negative leader produces unaccountably many VHF pulses (but is known to produce a single current pulse along the leader core) \citep{Hare:2020}. Furthermore, VHF emission has been detected from an upward jet (a type of upward lightning that penetrates above the thundercloud) which should only consist of streamers and not have a hot channel core \citep{Boggs:2022}. More detailed plasma physics, however, reveals a significant problem. Streamers evolve smoothly on a microsecond time scale, and thus do not naturally emit VHF \citep{Nijdam:2020}. The conclusion is that there must be an additional process that forces streamers to oscillate at nanosecond time-scales and thus emit VHF. This additional process is currently not known. Streamer models have suggested a few possible mechanisms. One is that when they are first created streamers grow at an exponential rate. This exponential growth could emit significant energy in lower VHF, but much less at upper VHF, where the exact cutoff depends on detailed model parameters \citep{Shi:2019, Malla:2024}. Another possible mechanism is streamer collision/merging. That is, when two oppositely charged and counter-propagating streamers encounter each other there will be a short surge in current between the two to equalize their potential, which is very efficient at emitting across the entire VHF band \citep{Shi:2019, Malla:2024}. Finally, a third mechanism was recently proposed that is a consequence of photo-ionization from a small number of photons during streamer propagation. The result is rapid Poisson fluctuations in streamer propagation that could emit across the entire VHF spectrum, but would have a very different spatial and temporal distribution from streamer collisions \citep{Malla:2024}. 
Figure \ref{fig:ModelSpectra} shows expected frequency vs time for these three models. All of these streamer models, however, have a serious difficulty that they only model one or two streamers simultaneously, where lightning leaders possibly have uncountable number of streamers at their tip \citep{Saba:2022}. It is thus possible, if not likely, that the VHF emission mechanism is an emergent phenomena from a large number of interacting streamers; which is beyond our current modeling capabilities. 

It is clear that high-bandwidth observations combined with high accuracy and sensitivity will be required to discern the mechanism behind lightning VHF emission. Most lightning observations thus far have focused on the lower VHF (30-100 MHz), leaving the upper VHF nearly entirely unexplored. One of the very few broadband lightning studies \citep{Pu:2021} did explore the spectrum of radio emission from lightning across the full VHF band, and reported that the VHF emission from positive leaders cuts off above 80~MHz, but negative leaders have higher frequency emission. Interpreting this through previous streamer models \citep{Liu:2019, Liu:2020}, the authors concluded that they could measure the electric field in the thunderstorm. SKA-LOW will be able to provide precise measurements across the entire VHF spectrum, and thus reveal tremendous insight into lightning plasma physics that cannot be obtained in any other way. \change{In addition to spectra, models are significantly further constrained by high-precision location information, which allow us to measure parameters such as width, speed, spreading angles, repetition rates etc... Performing such analyses across} different lightning stages (e.g. initiation, propagation, connection to ground, etc...) \change{we will be able explore the distribution of emission mechanisms during each process}, which is likely to change for different lightning stages. The VHF emission mechanisms, in turn, will give un-paralleled access to the plasma physics. For example, if the dominant mechanism is streamer collision then there must be counter-propagating oppositely charged streamers in the plasma.

\subsection{Imaging extremely quiet processes}

While some lightning processes are extremely loud (namely negative leader propagation), many lightning processes are very VHF quiet, in particular initiation and positive leader propagation. How lightning initiates is the top question in lightning science \citep{Dwyer:2014}. Lightning initiation is so VHF quiet that at this time, very few ``normal'' lightning initiations have been observed, by the LOFAR radio telescope \citep{Sterpka:2021}. There are also uncommon cases where lightning flashes are initiated by a VHF-loud process dubbed fast-breakdown \citep{Rison:2016}. However, this process is the exception, and will be discussed further below. The initiation event observed by LOFAR seems to smoothly increase in intensity from below the noise background of the telescope. Therefore, even LOFAR cannot observe the first streamers in a lightning flash, and it is entirely unknown if there is a significant length of time or not between the very first streamers in the lightning and the first process that LOFAR observed. Thus, it is unclear if LOFAR is truly observing the majority of the initiation process, or perhaps only the last stage of the initiation process, which makes it challenging to interpret the results. In addition, one would expect there to be numerous failed lightning initiations in the thundercloud. That is, bursts of streamers that do not develop into a hot plasma channel. However, this process has not been observed. Combined with the fact that we cannot observe the first streamers in a normal initiation with even LOFAR, the conclusion is that these failed lightning initiations likely do exist but are too VHF-weak for current instruments. Since we do not understand how streamers behave in thunderstorm conditions, or how they emit VHF in the first place, it is impossible to predict how intense these processes are in VHF.

Aside from initiation, the propagation of positively charged leaders is also very VHF-quiet, to the point that they have not been observed by LOFAR (despite knowing their exact location due to ancillary phenomena) \citep{Hare:2019, Scholten:2023b}. This is very difficult to understand since it is thought that streamers are the source of VHF emission from lightning, and positive leaders have plenty of streamers at their tips. It is very likely that the propagation speed of the positive leaders is an important factor, as positive leaders have been observed in VHF near ground \citep{Kong:2008, Pu:2021} and once in the cloud by the LWA (Long Wavelength Array) when a positive leader was propagating at a particularly high speed \citep{Stock:2023}. Combined with the fact that intracloud leaders are very hard to observe in optical (due to obscuring clouds), intracloud positive leader propagation is practically un-studied. 

SKA-LOW is the most sensitive instrument ever constructed in its frequency band. 
%Given the current push for building future interferometers in space, SKA-LOW may be the only chance we have for observing weakly emitting lightning processes.
Being able to observe both successful and failed lightning initiations will not only give us information about the lightning plasma physics (such as how easy or hard it is for streamers to turn into a hot conducting channel), but also about thunderstorm electrification. For example, how large of a region in a thunderstorm has a strong enough electric field to support electrical activity? How quickly can a thunderstorm re-charge after a lightning flash? Similarly, SKA-LOW is our best chance to regularly observe positive leaders and learn the clues needed to understand how lightning grows and emits VHF radiation. For example, how VHF emission is related to propagation speed, what is the spatio-temporal structure of the VHF emissions (e.g. how wide is the emitting region, or does it fluctuate in time), and how does positive leader VHF emission differ from VHF emission from negative leaders?

\subsection{Probing different kinds of thunderstorms with high precision}

The final case for SKA-LOW lightning observations is simply one of location. The only other instrument of comparable quality (LOFAR) is situated in the Netherlands, which is not known for energetic weather. Different climates produce different kinds of thunderstorms (in terms of altitude, energy, area, and distribution of charge). Dutch thunderstorms are particularly low-altitude and not-energetic (as compared to more equatorial thunderstorms for instance). For example, it is common in the USA for lightning flashes to reach 9-11~km altitude \citep{Edens:2014, Pu:2024}, where-as in the Netherlands it is uncommon for lightning flashes to exceed 7~km altitude \citep{Scholten:2021-HANL}. In addition, we have recently learned that the main positive charge pocket in Dutch thunderstorms is at the bottom of the cloud while thunderstorms elsewhere tend to have their main positive charge pockets near the top of the cloud \citep{Trinh:2025}. The result is that the Dutch climate significantly limits the lightning science that can be done with LOFAR, as there are many fascinating lightning phenomena that occur very rarely or not at all above the Netherlands. 

\begin{figure}
\centering
	\includegraphics[width=0.95\linewidth]{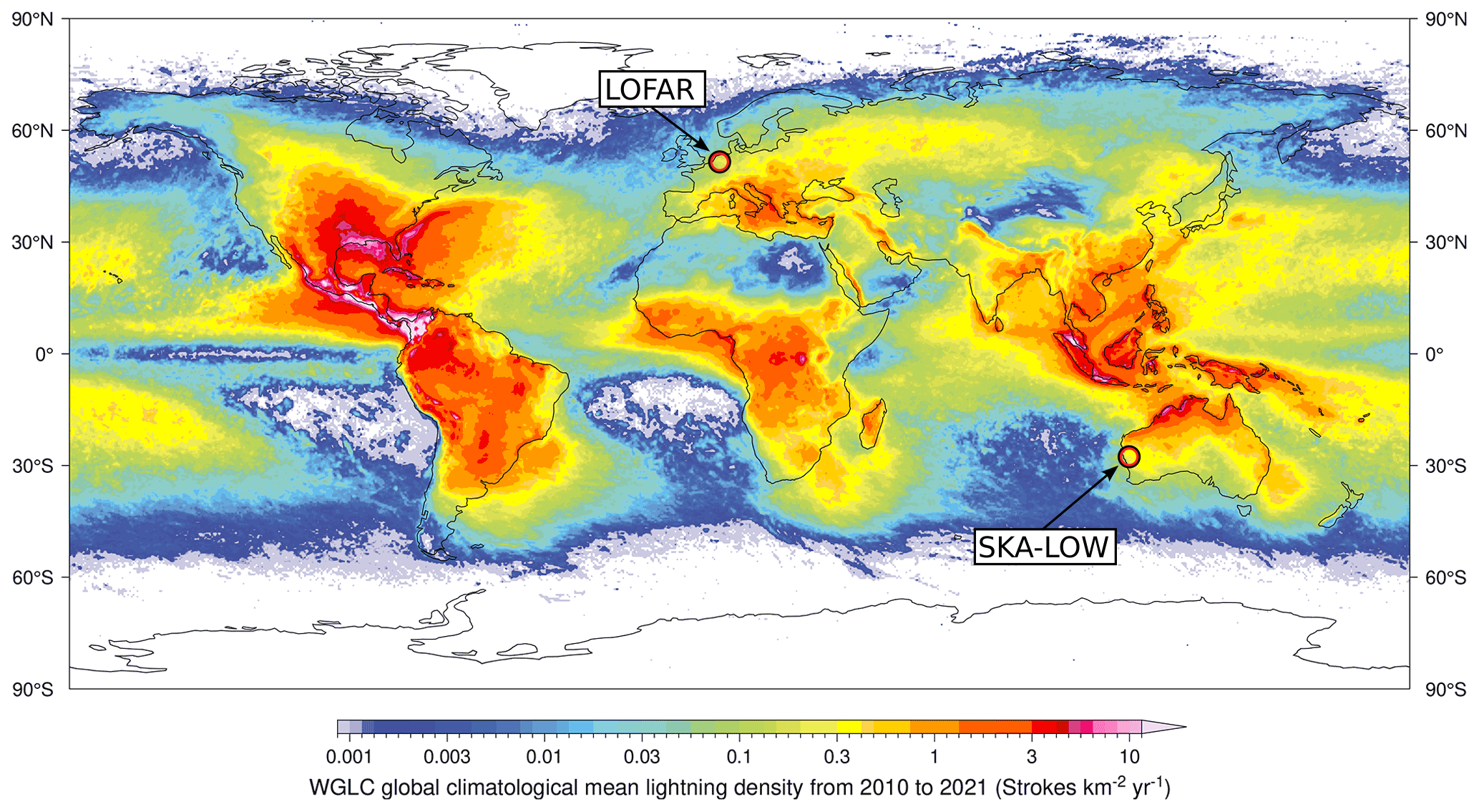}
	\caption{\label{fig:lightning_distribution}  Global distribution of lightning based on WWLN (world wide lightning network) data. Figure modified from  \citep{Kaplan:2022} figure 2. Color scale is logrithmic. }
\end{figure}

Since SKA-LOW, in western Australia, is in the middle of the desert that is much warmer than the Netherlands, it will be able to observe much more energetic lightning. Figure \ref{fig:lightning_distribution} shows a world distribution of lightning flashes as observed from space, and demonstrates that the Netherlands has about 1.5~flashes~km$^{-2}$~year$^{-1}$ , while the region around SKA-LOW has a lightning density of about 3~flashes~km$^{-2}$~year$^{-1}$ \citep{Kaplan:2022}. The exact flash density is not critical, but as a proxy of energy; demonstrating that the thunderstorms in western Australia are fundamentally different from those in the Netherlands. One specific highly-debated lightning phenomena is fast breakdown, which emits very strong VHF energy (but it is possible there are weakly-emitting examples) and propagates very quickly ($10^7$~m/s) \citep{Rison:2016, Tilles:2019, Pu:2022} (among many others). Fast breakdown seems to associated with the initiation of some lightning flashes \citep{Rison:2016}. The current hypothesis is that fast breakdown is a ``wave'' of streamers that propagates in a very strong electric field without a hot conducting channel core, however this hypothesis is difficult to test \citep{Liu:2019}. Fast breakdown is extremely fascinating as it is a strong VHF emitter, fast, associated with lightning initiation, and likely has fundamentally different propagation physics than other lightning phenomena. The LOFAR radio telescope, unfortunately, has never observed fast breakdown and it is not clear if this lack is due to location or other factors. Thus, SKA-LOW is likely the only opportunity we have to observe fast breakdown with the bandwidth, precision, and sensitivity needed to actually explore why fast breakdown occurs. In addition, there are many phenomena that happen more often in intense thunderstorms and thus should be significantly easier to explore with SKA-LOW than with LOFAR simply due to location; such as transient luminous events \citep{Dwyer:2014} and blues/sparkles \citep{Chanrion:2017, Scholten:2023}. 

\section{ SKA-LOW Lightning Observation and Imaging Strategy and Technical Requirements}

Lightning observations with SKA-LOW share many similarities with cosmic ray observations \citep{Huege01.2026.SKA}. Since lightning occurs in SKA-LOW's near field, and lightning processes tend to occur much faster than the integration time inherent to channelization, normal astronomical pipelines and imaging modes will not function for lightning science. Instead, similar to cosmic ray observations, we need to dump the raw-voltage buffers before channelization and post-process the voltages into scientific data products. We need a trigger signal to indicate when to freeze the raw-voltage buffers and read them to disk. Two possibilities include a self-trigger that simply watches for strong radio pulses observed by a SKA-LOW antenna, or a separate low frequency antenna placed in the vicinity of SKA-LOW ($\approx$10~km). Once a trigger signal is received, we will save out a second or two of raw-voltages from a few antennas from every station. 
We expect that SKA-LOW roll-out AA* will already provide the necessary baselines and sensitivity needed to perform extremely novel lightning science, however we also expect SKA-LOW AA4 to significantly improve the sensitivity and thus the extremely quiet lightning processes that can be observed.  

In post-processing we rely on two imaging techniques, originally developed for use with LOFAR. First we use our ``impulsive imager'' which is a sophisticated form of time-of-arrival \citep{Scholten:2021-init}. It is computationally cheap and easily images entire lightning flashes with high-precision. For example, figure \ref{fig:LofarFlash} was produced with this impulsive imager. Later, if we need to image a small section of a lightning flash with high detail, for example, the initiation or propagation of a lightning flash, we image that small region of a lightning flash with our near-field 3D beamforming imager: Time Resolved Imager 3D (TRID) \citep{Scholten:2022}. This imager relies fundamentally on beamforming in order to allow for extremely short (	$\approx$100~ns) integration times, and is capable of extracting the location of VHF source locations with 	$\approx$10~cm accuracy, as well as the full 3D emitting dipole orientations.

\section{Conclusion}

%Based on our experience with LOFAR, we propose to use

We will use the SKA-LOW telescope to image lightning with extremely high sensitivity and bandwidth while maintaining high-resolution. We have specifically focused on how SKA-LOW can be used to explore the source of VHF radiation from lightning, very weakly-emitting lightning processes, and probing a wider variety of lightning phenomena. However, there are many regimes of lightning science that will be made accessible by SKA-LOW. The broad goal is to understand how lightning initiates and propagates through the thunderclouds by imaging these processes with high enough spectral, temporal, and spatial resolution that the observations can be directly compared to modeling. 
%It should be noted that SKA-LOW may be our last chance to obtain this kind of high-resolution data of lightning, as future large radio interferometers will likely be built in space, and thus not capable of imaging terrestrial lightning with any accuracy. 
Since our observations will occur during thunderstorms that pollute normal astronomical data, our project uses observation time that is unusable to other science cases, thus is relatively resource cheap. Finally, studying lightning will highly benefit the SKA community as lightning data can be used to independently check time calibrations and antenna models, in addition, lightning captures the imagination of nearly every person on the planet and thus has strong potential for public outreach.

\bibliographystyle{abbrvnat-maxbibnames4}
\bibliography{references.bib} % if your bibtex file is called example.bib

\end{document}